\documentstyle[prl,aps,epsf,multicol]{revtex}

\begin{document}
\draft
\title{Anderson localization due to random magnetic field
in two dimensions
} 
\author{Akira Furusaki}
\address{
Yukawa Institute for Theoretical Physics, Kyoto University, Kyoto
606-8502, Japan\\
and Department of Physics, Stanford University, Stanford, California
94305} 
\date{\today}
\maketitle
\begin{abstract}
Results of large-scale numerical simulations are reported on the
Anderson localization in a two-dimensional square lattice
tight-binding model with random flux.
Localization lengths, fluctuations of the conductance, and the density
of states are computed for quasi-one-dimensional geometry.
Numerical results indicate that the model exhibits the same critical
behavior as the one studied by Gade and Wegner.
It is argued that all the states except a zero-energy state are
localized and the density of states has a singularity in the center of
the band. 
The energy scale below which the density of states increases is found
to be extremely small ($\lesssim10^{-2}$).
\end{abstract}
\pacs{71.30.+h,72.10.Bg,71.55.Jv}

\begin{multicols}{2}

It is general wisdom that noninteracting electrons are localized in
two-dimensional (2D) disordered systems\cite{Abrahams}.
There are, however, some well-known exceptions to this rule.
These include electrons having strong spin-orbit
coupling\cite{HLN} and integer quantum Hall systems\cite{Huckestein}.
Recent studies have shown that 2D Dirac fermions with random
gauge field offer another exception to the rule\cite{Ludwig,Mudry}.
For a model of 2D nonrelativistic fermions subjected to random
magnetic field with zero mean, the existence of delocalized
states has been a subject of debate.

The random flux model, in which static magnetic field
is randomly distributed with zero mean, got much attention recently in
connection with the gauge field theory of high-$T_c$ superconductivity
\cite{Nagaosa} and the composite-fermion theory of the half-filled
Landau level \cite{HLR}. 
It has been controversial, however, whether this model has a delocalized
state \cite{LeeFisher}.
On the one hand, several numerical and analytical studies concluded that
all the states are localized and belong to the unitary class of the
scaling theory\cite{Sugiyama,DKKLee,Aronov,YBKim,Yakubo,Batsch}.
On the other hand, a different conclusion that there are delocalized
states near the center of the band was reached by other
people\cite{Avishai,Kalmeyer,Zhang,Liu,Sheng,Miller,KunYang,Xie}.
One source of the discrepancy in numerical works is the extremely large
localization length near the band center, making it difficult to decide
whether or not states are localized from numerical data of finite-size
systems. 

In this paper I present various numerical results obtained through
the largest numerical simulations performed so far for the square
lattice tight-binding model subjected to random flux with zero mean.
The results indicate that a state at the band center ($E=0$) is not
localized.
This is reminiscent of the integer quantum Hall system.
There is, however, a crucial difference:
the density of states (DOS) is found to be divergent at $E=0$ in
the random flux case.
This behavior is similar to the 1D and 2D random hopping models
\cite{Cohen,Eilmes},
and a crucial role is played by a special particle-hole symmetry
relating a state of energy $E$ with a state of energy $-E$.
The random flux model is argued to be in the same universality class
as a model studied by Gade and Wegner \cite{Gade}.
Although this was already anticipated in \cite{Ludwig,Miller},
this Letter reports for the first time that the random flux model
shares a hallmark of the Gade-Wegner model, i.e., the divergence of
the DOS at $E=0$. 

\begin{figure}
\begin{center}
\leavevmode
\epsfxsize=80mm
\epsfbox{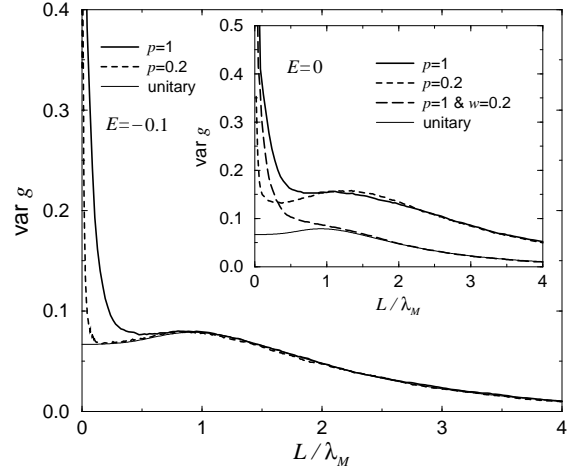}
\end{center}
\caption{Variance of $g$ as a function of the length of the
disordered region at $E=-0.1$.
${\rm var}\,g$ approaches 0 for $L$ shorter than
mean free path, although invisible in this scale.
Inset: Variance of $g$ at $E=0$.
The thin curves are the analytic result for the unitary ensemble
{\protect\cite{Mirlin}}, where var$g\to1/15$ as $L\to0$
(with $M\ll L$).
For both figures $M=32$.
}
\label{fig:varg}
\end{figure}

The Hamiltonian of the tight-binding model is
\begin{equation}
H=
-\sum_j\sum^M_{k=1}
\left(c^\dagger_{j+1,k}c_{j,k}
      +e^{i\theta_{j,k}}c^\dagger_{j,k+1}c_{j,k}
      +{\rm H.c.}\right),
\label{H}
\end{equation}
where $c_{j,k}$ is annihilation operator of a fermion on site
$(j,k)$.
The random magnetic flux is introduced through the random Peierls
phase $\theta_{j,k}$ in the hopping matrix element.
The magnetic flux
$\phi_{j,k}=\theta_{j,k}-\theta_{j-1,k}$
takes a random number in $-\pi p\le\phi_{j,k}\le\pi p$
with a uniform distribution.
The parameter $p$ is set to be 1, except in Fig.~\ref{fig:varg}.
Numerical calculations are done for samples that have quasi-1D geometry of
width $M$ in the $y$ direction and of
length $L$ in the $x$ direction ($M\ll L$).
Periodic boundary condition is imposed in the $y$ direction
($c_{j,M+1}=c_{j,1}$), whereas open boundary conditions are assumed in
the $x$ direction for most of the calculation.
For even $M$ the lattice can be divided into A and B sublattices.
For each eigenfunction $\psi_E$ with energy $E$, changing sign of
$\psi_E$ on every site of, say, the A sublattice yields a new
eigenfunction $\psi_{-E}$ with energy $-E$\cite{Miller,Ohtsuki}.
This symmetry relating the $\pm E$ states holds
for each disorder configuration.
For odd $M$, however, the particle-hole symmetry is absent under the
periodic boundary condition.

\begin{figure}
\narrowtext
\begin{center}
\leavevmode
\epsfxsize=92mm
\epsfbox{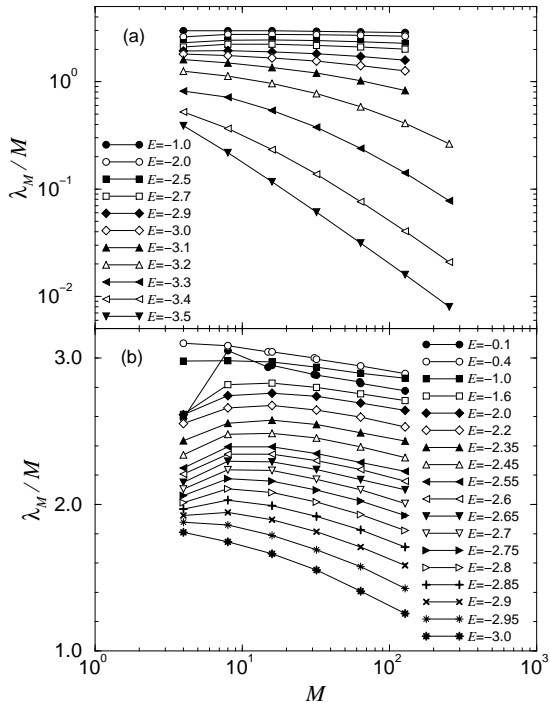}
\end{center}
\caption{Localization length for quasi-1D geometry calculated for
$M=4$, 8, 16, 32, 64, 128, and 256.  Additional data of $M=15$ and
31 are shown for $E=-0.4$ and $-1.0$.
In Fig.~{\protect\ref{fig:loclength}}(b) the statistical error for the
data at $M=128$ is about the same size as the symbols, whereas for
smaller $M$ the error bar is much smaller than the symbols.
The kink at small $M$ in the $E=-0.1$ data should be finite-size
effects.} 
\label{fig:loclength}
\end{figure}

The localization length is calculated from the
exponential decay of the retarded Green's function obtained by using
the standard recursive algorithm \cite{MacKinnon}:
$\langle\ln\|G^r_E(1,k;L,k')\|\rangle\sim-L/\lambda_M$, where $\|G\|$
and $\langle\ \rangle$ denote norm of $G$ and ensemble average,
respectively.
Figure \ref{fig:loclength} shows the quasi-1D localization length
$\lambda_M$ normalized by $M$ as functions of $M$ and $E$.
The typical length of quasi-1D samples used in the calculation is
$3\times10^5$, $4\times10^5$, and $8\times10^5$ for $M=32$, 64, and 128,
respectively.
Furthermore, ensemble average is taken, typically, over 70 ($20$)
samples for $M\le64$ ($M=128$) to reduce the statistical error.
The quality of the numerical data is therefore greatly improved from the
earlier numerical results \cite{Sugiyama,Yakubo,Kalmeyer,Liu,Miller}.
Clearly the states near the band edges ($|E|>3.0$) are localized
[Fig.\ \ref{fig:loclength}(a)].
Figure \ref{fig:loclength}(b) shows $\lambda_M/M$ decreases
as $M$ increases, suggesting that the states with $|E|\ge0.1$ are
all localized in the 2D limit.

\begin{figure}
\begin{center}
\leavevmode
\epsfxsize=80mm
\epsfbox{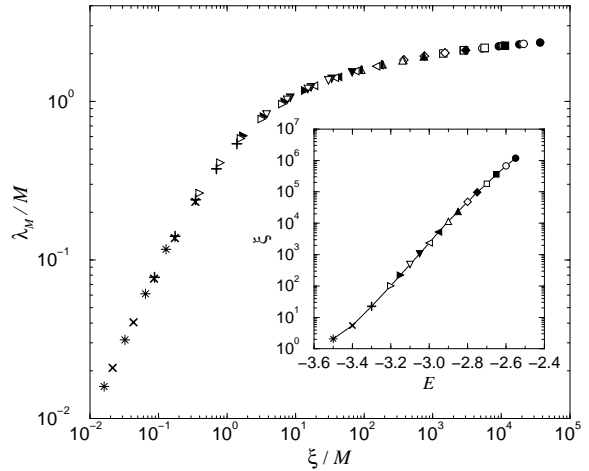}
\end{center}
\caption{Scaling curve obtained from the data for $-3.5\le
  E\le-2.55$. 
For $-3.0<E\le-2.55$ only the data of $M\ge32$ are used.
Inset: Localization length versus energy.}
\label{fig:scaling}
\end{figure}

The localization lengths of the quasi-1D wires are expected to
satisfy the one-parameter scaling $\lambda_M/M=f(\xi/M)$,
where $\xi$ is the localization length in 2D.
The scaling indeed holds as shown in
Fig.~\ref{fig:scaling}\cite{note}. 
The scaling curve quantitatively agrees with the earlier results of
Refs.\ \cite{Sugiyama} and \cite{DKKLee}.
The agreement with the latter work is somewhat surprising in that
a network model is used in \cite{DKKLee} which is an effective
model in the semiclassical limit.
The 2D localization length $\xi$ grows exponentially and reaches
$10^6$ lattice spacings at $E=-2.55$; see inset.

\begin{figure}
\begin{center}
\leavevmode
\epsfxsize=80mm
\epsfbox{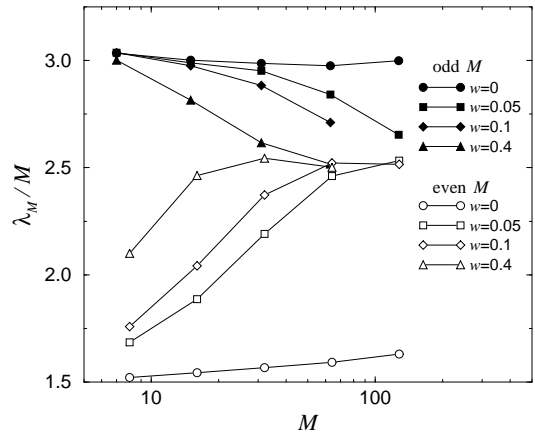}
\end{center}
\caption{Localization length at $E=0$ as functions of $M$ and
the on-site disorder $w$.
The filled symbols represent data for $M=7$, 15, 31, 63, and 127.
The open symbols are data for $M=8$, 16, 32, 64, and 128. }
\label{fig:onsite}
\end{figure}

Figure \ref{fig:onsite} shows
$\lambda_M/M$ versus $M$ at $E=0$ \cite{note2}.
There is a striking even-odd effect in the $E=0$ data, as
noticed earlier in Refs.\ \cite{Sugiyama2,Miller}.
A new finding here is that $\lambda_M/M\bigr|_{E=0}$ stays almost
constant for odd $M$ while it gradually increases for even $M$,
suggesting that $\lambda_M/M\to{\rm const}(>0)$ as $M\to\infty$.
This would mean that $\psi_{E=0}$ is a critical or multifractal wave
function, as suggested by Miller and Wang\cite{Miller}.
By contrast, at $|E|=0.1$, there is little even-odd oscillations
[Fig.~\ref{fig:loclength}(b)], and $\lambda_M/M$ is a decreasing
function of $M$.
The importance of the particle-hole symmetry can be also seen by
examining the effects of on-site disorder, which breaks the symmetry.
The on-site disorder is introduced by adding a term
$\sum_{j,k}\epsilon_{j,k}c^\dagger_{j,k}c_{j,k}$ to $H$,
where $\epsilon_{j,k}$ are taken to be randomly distributed in the
interval $[-w/2,w/2]$.
Figure \ref{fig:onsite} clearly shows that in the presence of the
on-site disorder $\lambda_M$'s of even and odd $M$'s merge together
and decrease for $M>M_c$.
The crossover width is $M_c \approx 64$ (300) for
$w=0.4$ (0.05) and would diverge as $w\to0$.
These results strongly suggest that the wave function
is critical in the 2D limit only when $E=w=0$, and that small $|E|$
or $w$ is sufficient to change it to a localized one.

The states away from the band center belong to the unitary class.
This can be verified by calculating fluctuations of two-terminal
conductance as a function of $L$.
For this purpose, perfect leads are attached to both
ends of quasi-1D wires, and the transmission matrix $t$ is
calculated from the Green's function $G^r_E$.
The dimensionless conductance $g$ is then obtained from the Landauer
formula, $g={\rm Tr}(tt^\dagger)$.
Figure \ref{fig:varg} shows
${\rm var}\,g=\langle g^2\rangle-\langle g\rangle^2$ for $M=32$,
averaged over $2\times10^4$ samples.
For $|E|=0.1$, ${\rm var}\,g$ is calculated for $p=1$ and $0.2$ without
the on-site disorder.
Almost identical ${\rm var}\,g$ versus $L/\lambda_M$ curves are obtained
for $|E|=1.0$ and 0.02 as well.
A thin line in Fig.~\ref{fig:varg} shows ${\rm var}\,g$ of the unitary
ensemble calculated in the limit $M\ll L$
by Mirlin {\it et al.}\cite{Mirlin} using
the supersymmetric $\sigma$ model approach.
Notice that, except for the peaks at $L<0.5\lambda_M$, the numerical
results of $|E|=0.1$ are indistinguishable from the thin line (unitary
ensemble).
The discrepancy occurs only for $L\lesssim M$, where the samples are no
longer quasi one-dimensional.
The numerical curve of $p=0.2$ is closer to the analytic result because
$M/\lambda_M|_{p=0.2}\ll M/\lambda_M|_{p=1}$.
These results clearly show that for $|E|\ge0.1$ and $p\ge0.2$ the wave
functions belong to the unitary class.

The variance of $g$ has a different $L/\lambda_M$-dependence at $E=0$
for even $M$; see inset of Fig.~\ref{fig:varg}.
Without the onsite-disorder, for each $L/\lambda_M$, ${\rm var}\,g$
of $E=0$ is larger than ${\rm var}\,g$ of $E\ne0$ \cite{Ohtsuki2}.
This clearly shows that, when $w=0$, the zero-energy state does not
belong to the unitary class.
The on-site disorder, however, drives $\psi_{E=0}$ back to the unitary
class, as shown by the long-dashed line ($p=1$ \& $w=0.2$) in the
inset. 
These observations are consistent with Fig.~\ref{fig:onsite}.

It seems quite natural to assume that the states belonging to the
unitary class in the quasi-1D geometry remain to be in the same class
as $M\to\infty$.
This would mean that all the states away from the band
center are localized.
A state at $E=0$, if it exists, should not be localized in 2D.
It follows both from the recent result \cite{note2} that the state at
$E=0$ is delocalized for odd $M$ under open boundary conditions in
the $y$ direction and from the numerical data in
Fig.~\ref{fig:onsite}.
The delocalization of the zero-energy state is inferred by requiring
that the 2D limit ($M\to\infty$) should be independent of the
boundary conditions \cite{note3} and of the parity of $M$.
The delocalization at the band center is a consequence of the
particle-hole symmetry as in the random hopping model \cite{Cohen} and
will be ruined by the onsite-disorder.

\begin{figure}
\begin{center}
\leavevmode
\epsfxsize=80mm
\epsfbox{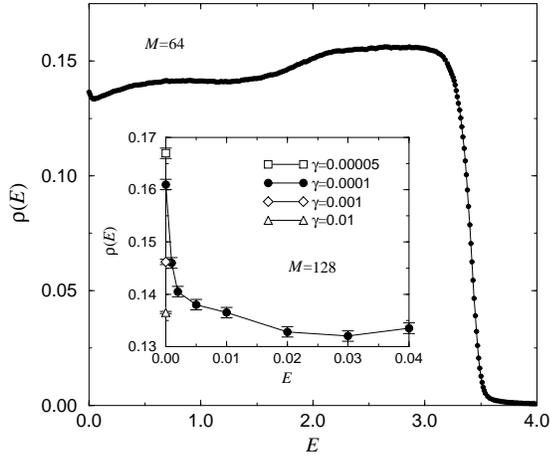}
\end{center}
\caption{
Density of states of a system of $M=64$ and $L=128000$ calculated with
$\gamma=10^{-2}$.
Inset: $\rho(E)$ of a system of $M=128$ and $L=64000$, 64000, 204800,
and 256000 for $\gamma=10^{-2}$, $10^{-3}$, $10^{-4}$, and
$5\times10^{-5}$, respectively.
}
\label{fig:dos}
\end{figure}

As pointed out in \cite{Ludwig,Miller}, the random-flux model has the
same symmetry property as the Gade-Wegner model, and it is natural to
expect that the two models share the same critical behavior.
In the Gade-Wegner model the localization length diverges towards the
band center, where the DOS $\rho(E)$ is also divergent as
$\rho(E)\sim\exp(-c\sqrt{\ln|1/E|})/|E|$ ($c$: constant)\cite{Gade}.
The characteristic energy scale below which the singularity of the DOS
manifests itself is then $E_c=\exp(-c^2)$, which can be
extremely small depending on $c$.
This may explain why no singularity was found in $\rho(E)$
before \cite{Sugiyama,Miller,Ohtsuki,Verges}. 
To find the presumably weak singularity, I computed the DOS with
high accuracy using the recursive method \cite{MacKinnon2}.
In this calculation a small imaginary number was added to the energy
($E\to E+i\gamma$), instead of attaching perfect wires.
This amounts to averaging $\rho(E)$ over the energy interval of order
$\gamma$.
Figure \ref{fig:dos} shows the DOS of a system of $L=128000$ and
$M=64$ with $\gamma=10^{-2}$.
The overall shape of the DOS is similar to the one obtained by the
retraced-path approximation \cite{Gavazzi}.
Notice, however, the tiny peak centered at $E=0$.
Its height grows with smaller $\gamma$ and larger $M$ (inset),
which is a clear signature of the divergent DOS.
To determine the precise form of the singularity requires further
investigation. 
It is important to note here that $\gamma$ is
kept large enough to smear out the microscopic structure
in $\rho(E)$ near $E=0$.
Because of the level repulsion and of the particle-hole symmetry,
$\rho(E)$ vanishes at $E=0$ for even $M$ \cite{Slevin}.
It is expected that, in the limit $M\to\infty$, the dip in the DOS at
$E=0$ disappears and $\rho(E)$ diverges at $E=0^+$, in analogy with
the 1D random-hopping model with even number of sites\cite{Cohen}. 
The moderate smearing due to $\gamma$ helps revealing the diverging
behavior.

The discovery of the divergent DOS at $E=0$ establishes the connection
between the lattice random flux model and the Gade-Wegner model.
The critical behavior of the latter model is closely related to
the model of Dirac fermions with random gauge
field\cite{Ludwig,Mudry},
which has the same particle-hole symmetry.
This supports the conclusion based on the symmetry argument
that a state at the band center is the only delocalized state
for any $p$ ($0<p\le1$) in the absence of the on-site disorder.
In models without the particle-hole symmetry, all
the states should be localized, in agreement with
\cite{DKKLee,Aronov,YBKim,Batsch}. 

I have greatly benefited from discussions with A.\ G.\ Abanov,
P.\ W.\ Brouwer, R.\ Gade, D.\ K.\ K.\ Lee, P.\ A.\ Lee, C.\ Mudry,
N.\ Nagaosa, C.\ Nayak, N.\ Taniguchi, and X.-G.\ Wen.  
I am grateful to the condensed-matter theory group at Stanford for
hospitality, where this work was completed, and to D.\ K.\ K.\ Lee for
helpful comments on the manuscript. 
This work was supported by a Grant-in-Aid for Scientific Research
(No.\ 09740278) and by a grant for overseas research, both from the
Ministry of Education, Science and Culture of Japan. 
The numerical computations were carried out at the YITP
Computing Facility and at the Supercomputer Center of ISSP,
University of Tokyo.

\end{multicols}


\begin{references}
\bibitem{note} The data of $|E|<2.55$ are not used to make the scaling
 plot because of their larger statistical error.
\bibitem{Ohtsuki2} The enhancement of ${\rm var}\,g$ at $E=0$ was
 noticed in \protect\cite{Ohtsuki}.
\bibitem{note3} The boundary conditions are changed by the boundary
 operator, $\sum_j(e^{i\theta_{j,M}}c^\dagger_{j,1}c_{j,M}+{\rm
 H.c.})$, which should not affect bulk properties in the limit
 $M\to\infty$.
\end{references}
\end{document}